# Business Models for e-Health: Evidence from Ten Case Studies

Chris Kimble

**Abstract**
An increasingly aging population and spiraling healthcare costs have made the search for financially viable healthcare models an imperative of this century. The careful and creative application of information technology can play a significant role in meeting that challenge. Valuable lessons can be learned from an analysis of ten innovative telemedicine and e-health initiatives. Having proven their effectiveness in addressing a variety of medical needs, they have progressed beyond small-scale implementations to become an established part of healthcare delivery systems around the world.

## 1   Introduction

A cursory glance at the popular press leaves no doubt that we are in the midst of a healthcare crisis. Costs are spiraling out of control, partly because of the increased burden of managing chronic diseases in an aging population, and partly because of the cost of medical advances that make this possible.

A report from Standard & Poor (Mrsnik & Morozov, 2012) notes that, for the majority of the G-20 economies, expenditure on healthcare will be the fastest growing item in their budgets over the coming decades. It adds that healthcare already accounts for more than 60 percent of the increase in age-related spending, and predicts that Germany, the United States, the United Kingdom, France, and other nations face increases in excess of 6 percent of their GDP. The report also identifies non-demographic factors, such as the cost of new treatments and technologies, as contributing to rising costs. Hill and Powell (2009) concur, observing that, in the United States, "accelerating costs are a natural outcome of excess demand for medical treatment resources and greater complexity in the art of medicine" (Hill & Powell, 2009, p. 266).

The European Union (EU) faces similar problems. Although each of the 28 member states has a different method of delivering and charging for healthcare, EU Directive 2011/24/EU set out to facilitate cross-border healthcare and patient mobility within Europe, adding an additional layer of cost and complexity to the provision of medical services. Thus, in addition to the challenge of dealing with an aging population, the EU also faces the administrative, logistic, and political problems of a proposed pan-European system of healthcare founded on services provided and funded by independent nation states (Currie & Seddon, 2014).

One solution put forward to address these problems is to use information technology (IT) to streamline healthcare delivery so that it becomes more cost-efficient (Kimble, 2014). Yet, technology in and of itself is not a solution. As Chesbrough notes, "The economic value of a technology remains latent until it is commercialized in some way" (Chesbrough, 2010, p. 354). Gamble et al. (2004) observe that while the clinical effectiveness of applications, such as telemedicine, is generally acknowledged, there remains considerable controversy about how to assess the costs and benefits. Without a self-sustaining business model, such innovations risk becoming part of the problem rather than part of the solution.

## 2   Information Technology and Healthcare

There is no doubt that modern healthcare is an information-intensive activity that, superficially at least, would appear to be ripe for computerization. For example, Davenport and Glaser (2002) indicate that a typical physician will need to know something about 10,000 diseases, 3,000 medications, and 1,100 laboratory tests, and to keep up to date with the 400,000 articles that are added to the biomedical literature each year. In addition, the provision of modern healthcare requires the coordination of activities across a wide range of primary healthcare providers, such as acute-care hospitals, hospices, and rehabilitation centers, as well as numerous other organizations that are not direct providers of healthcare, such as laboratories, insurance companies, and pharmacies.

Despite this, Heeks (2006) contends that most health information systems fail in some way, while Chen et al. (2013) maintain that up to 75 percent of e-health programs fail during their operational stage. Outside the realm of IT in medical technology—that is, when IT is used as part of some medical intervention — the application of IT to healthcare tends to fall under one of two broad categories: telemedicine and e-health.

Telemedicine, the older of the two terms, dates from the 1950s, when it was used to describe the transmission of radiology images via a standard telephone system (Zundel, 1996). Since then, numerous definitions have been put forward and various terms have been coined. Perhaps the simplest is that telemedicine is "a system of healthcare delivery in which physicians examine patients through the use of telecommunications technology" (Celler et al., 2003, p. 242), now more often referred to as a teleconsultation. Telemedicine is relatively well established in countries such as the United States and Australia, and there is growing interest in it among developing countries, where problems of transport and access to healthcare facilities make telemedicine an attractive proposition (Chen et al., 2013).

On the other hand e-health, "the application of information and communications technologies across the whole range of functions that affect the health sector" (EU, 2004, p. 4) is a more recent term that, according to Eysenbach (2001), first began to be used in 1999 and became more widely used after its adoption by the European Commission in 2004. This umbrella term is used to cover a wide range of services and systems that combine medicine/healthcare and IT, such as telemedicine, electronic health records, healthcare information systems, and certain aspects of telework, such as the operation of "virtual" healthcare teams. It is also sometimes used to include m-health, which is the use of wearable electronic monitoring systems.

Clearly, the two terms are related, as telemedicine could form part of an e-health initiative and e-health could equally be part of the support system for telemedicine. In the sections that follow, unless there is a specific reason for identifying one type of initiative or the other, e-health will cover both telemedicine and e-health.

## 3 Business Models

A business model is a high-level conceptual description of the activities of value creation (the creation of a product that meets customers' needs), value capture (the marketing, support, and sale of the product), and value architecture (the chain of activities that link customers to the suppliers of a product). Although some argue that business models have always existed, the term first began to be used during the "dot-com bubble" of the late 1990s, when it was to describe the business rationale behind the many web-based businesses that grew up at the time (Osterwalder et al., 2005). Chesbrough and Rosenbloom (2002) argue that this surge in popularity stemmed, in part, to the inability of other methods to deal with the range of new business opportunities that electronic commerce opened up.

The idea behind a business model is to create a simplified, abstract view of the way a company operates by reducing it to its essential elements and identifying the relationships among them, thus highlighting the firm's core business logic (Linder & Cantrell, 2000). Over time, the concept has evolved from being simply a term that refers to the logic of the firm or a way of doing business to a conceptual and analytical framework consisting of a fixed set of elements or building blocks (Kimble & Bourdon, 2013). There is now a variety of such frameworks that can be used to describe business models; the one used here is the business model canvas (Osterwalder & Pigneur, 2010).

The business model canvas is a template for developing new business models or documenting existing ones. It is designed for use by a group of people to either brainstorm a new business model, or to discuss, analyze, and change an existing business model. It splits a business model into four components — the client, the offer, resources, and finance — which are then further broken down into nine building blocks, each interacting with the other (see Exhibit 1).

| Component | Building Block |
|---|---|
| Client | Customer Segments |
| | Customer Relationships |
| | Value Delivery (Distribution Channels) |
| Offer | Value Proposition |
| Resources | Key Activities |
| | Key Resources |
| | Key Partnerships |
| Finance | Revenue Streams |
| | Cost Structure |

**Exhibit 1. The Business Model Canvas**

### 3.1 The Application of Business Models to e-Health

Chen et al. (2013) note that although sustainable business models need to be developed in order to create value for a healthcare provider as well as for the patient, research on business models for e-health is limited. They claim that a review of more than 500 peer-reviewed articles found only 38 that contained data on the costs and benefits of telemedicine. Gamble et al. (2004) attributes this to the fact that most articles tend to address the question, "Can we do this?" rather than "Should we do this?" They note that e-health raises particular challenges for the evaluation of business models: Systems may have multiple uses, and joint costs are difficult to apportion; the existence of a system in one part of an organization may lead to expanded use in another; and, as with most ventures connected to IT, technological change can rapidly make existing ideas outdated.

The approach here is to focus on e-health initiatives that are part of established systems that have already demonstrated their effectiveness. Like Heeks (2006) and Chen et al. (2013), Broens et al. (2007) observed that many e-health initiatives do not progress beyond pilot programs, and they set out to examine the factors that influence e-health initiatives. Following a literature review, they classified the determinants of success in telemedicine implementation (see Exhibit 2).

| Factor | Stage in life cycle |
|---|---|
| Technology | Experimental Prototype |
| Acceptance | Pilot Project |
| Organization | Small-Scale Use |
| Finance | Large-Scale Use |
| Legal and Social Policy | Operational Product |

**Exhibit 2. Factors that Influence the Implementation of Telemedicine**

They contend that this classification is, in effect, a layered implementation model in which the critical success factor changes according to the level of maturity of the implementation. For example, in the experimental prototype phase, technological issues, such as the availability, quality, and level of support for the technology, are most important. When small-scale pilot studies become larger in scope, however, finance and organizational issues become increasingly important. The implication is that finding the correct business model tends to become more important as projects increase in scale. Thus, if we are looking for examples of successful business models for e-health, the focus should be on large-scale projects.

## 4   Ten Case Studies Shed That Light on e-Health

Chen et al. (2013) review eight case studies: three from developed economies, and five from developing economies. In their report for the European Commission, Valeri et al. (2010) examine five case studies from Europe. Both teams of researchers use Osterwalder and Pigneur's (2010) business model canvas to describe the business models they found, which means that the two studies can be compared.

Although it may not seem to be immediately relevant to the problems outlined above, it is helpful to include case studies from developing economies. Not to do so would narrow the choice of successful business models unnecessarily. Just because a business model has been created within a developing economy does not mean that it cannot be applied elsewhere. Some even argue that business models forged in low-income economies can have certain advantages over those created in more affluent settings (Hart & Christensen, 2002). Exhibit 3 summarizes the ten case studies described below. Each is broken down into the four key components of offer, client, resources, and finance.

### 4.1   Apollo Telemedicine (India)

The Apollo Telemedicine Networking Foundation is a nonprofit organization started by the Apollo Hospitals Group, India's largest private healthcare organization. Its goal is to offer a successful working model of telemedicine for the developing world that is capable of providing continuous access to sophisticated medical support systems. It connects Apollo hospitals with centers located in rural areas, and is now India's largest telemedicine provider, with 150 centers in India and overseas providing teleconsultation and tele-education to 53 countries.

India is the second most populous country in the world after China. Although urban areas, such as Mumbai and Delhi, are growing rapidly, most people still live in rural settings and do not have ready access to hospitals or to medical specialists. Consequently, Apollo's primary offer is ready access to the specialized knowledge available in hospitals located in the large cities, without the need for the cost and disruption of travel. Apollo also makes some secondary value propositions by offering a secure and confidential service to its clients and building its brand image through its link with other countries via its tele-education program.

Apollo's clients are principally low- to middle-income individuals in rural India, and employees of various government organizations and private companies, for whom it provides healthcare. Much of the hardware, software, and satellite links used by the network are provided free of charge by IRSO, the Indian space research organization. Other equipment and the cost of healthcare centers are paid for via consultation fees, which range from $US 20 to $US 30 per consultation. Income from companies and government organizations provide a separate revenue stream. Apollo also makes money from managing patient records.

### 4.2 Aravind Tele-Ophthalmology (India)

Aravind is the largest eye-care organization in the world. Founded in 1976 in Madurai, India, with the aim of eliminating blindness, it established vision centers in rural villages. The centers are connected to higher-level hospitals through videoconferencing so that an ophthalmologist in the hospital can interact directly with a patient in a rural location. A local doctor works with the patient and uses dedicated ophthalmic equipment to transfer the images to the hospital for diagnosis.

Aravind's clients are low-income and marginalized individuals. As with Apollo, part of the offer is access to the specialized knowledge without the cost and disruption of travel, but the primary value proposition is free or affordable eye care to treat or avoid blindness for individuals who would otherwise have no access to a doctor.

A nonprofit organization, Aravind maintains a number of hospitals, vision centers, screening vans, and Internet kiosks, which are linked with a high-speed wireless network. Most of the personnel, including the ophthalmologists, work as volunteers. Paid consultations cost only $US 0.50, but 65 percent of consultations and 75 percent of surgical interventions are provided free of charge. There is, however, a small charge for transport to a hospital. Staff and operating costs are kept to an absolute minimum so that revenue and voluntary donations can be used to subsidize non-paying patients.

### 4.3 Arizona Telemedicine Program (United States)

The Arizona State Telemedicine Program was established in 1996 by the Arizona State legislature with a $1.13 million federal grant. It acts as a purchasing agent for member institutions' equipment and telecommunication infrastructure. Currently the Arizona Telemedicine Program helps provide medical services to 20 communities and educational material to 34 communities. It also provides support for the development of new telemedicine projects.

The value proposition for the Arizona telemedicine program differs from most others in that it aims to facilitate medical services rather than provide them. It acts as a broker helping to reduce the cost of developing and providing e-health networks by sharing the costs between institutions. It also provides training and consulting services. Its clients are for-profit and nonprofit organizations in Arizona, including prisons, public schools, and private hospitals.

The communications network is leased from utility companies and most aspects of the program are managed by staff at the telehealth business office of the University of Arizona. External experts provide specialized medical and legal advice. Most of the income is from grants made by the Arizona state government and other donors. In addition, clients pay an annual membership fee and are charged for each service they receive.

### 4.4 Centro Unico di Prenotazione (Italy)

The Centro Unico di Prenotazione is a system that allows patients to book, reschedule, cancel, and pay for visits to specialists or tests requested by their doctor. It serves patients in the Umbria region of central Italy, which has a population of almost 900,000 people, about half of which live in rural areas. The population is aging: 55 percent are over 50, and more than 70 percent of them live in rural areas. Initiated in 1999, the system now services many of Umbria's pharmacies, medical specialists, and laboratories.

The clients are the residents of Umbria, doctors in the region, and the regional health authorities. Before the introduction of the system, patients could only book, cancel, reschedule, and pay for tests and specialist visits at one of the two regional hospitals or at one of four regional health centers. Now this can be done at any of the local pharmacies or medical-surgical units that are part of the system. Similarly, tests and specialist visits can be arranged with any of the doctors or diagnostic centers in the region. The pharmacies that are part of the system benefit from increased income; the health authorities use the system to manage waiting lists more accurately and efficiently, and to monitor the effectiveness of their health awareness campaigns.

The system is based around a simple client-server application using off-the-shelf technology and a database maintained by the regional government, which also provides training and an IT help desk for system users. Payments to doctors, laboratories, and pharmacies are made via the system by the patients, who are subsidized by the Italian heath service. The key to the success of the initiative is the active involvement of a network of local doctors, laboratories, and pharmacies. To encourage participation, pharmacies are given an additional payment of 2 euros for each booking made through their system.

### 4.5    Myca Nutrition (Canada)

Myca Nutrition is a web and mobile communications platform created to connect nutritionists with their clients. A nutritionist can carry out a consultation via videoconferencing, instant messaging, telephone, secure email, or the nutritionist's own website. All the nutritionist needs is a computer and Internet access; the client needs a cell phone or a computer. The platform provides a billing service so that the nutritionists are reimbursed for their services.

As an application, Myca Nutrition is aimed primarily at nutritionists and doctors, although both doctors and patients use the system. For the nutritionists, Myca Nutrition provides a way to reach more clients and attract new referrals; it also provides a convenient way to manage payments. For the patients, it is a convenient way to get dietary advice and receive feedback.

The system uses standard web technology to provide an interactive web space with facilities for secure online billing and payments, and a database that clients can search to find a doctor or nutritionist that suits their needs. Both doctors and patients pay a monthly fee. Patients pay doctors directly through the Myca Nutrition website. Myca Nutrition also franchises it system to other countries.

### 4.6    People's Liberation Army Telemedicine Network (China)

The People's Liberation Army telemedicine network uses satellite and landlines to link 114 military hospitals, 97 civil hospitals, and more than 300 specialists across China. It is the oldest of three telemedicine networks in China; the others are the Golden Health Network and the International MedioNet of China. So far, 3560 teleconsultations have been conducted, including emergencies, and 60,000 people have participated in distance education sessions.

The situation in China, the most populous country in the world, is similar to that of India. There is a large gap between urban and rural areas in terms of the provision of healthcare. As with India's Apollo network, the primary value proposition here is ease of access to specialized knowledge, without the cost of travel, combined with the secondary value proposition of links to expertise from other countries via a tele-education program.

The People's Liberation Army Telemedicine Network is funded by the central government from its military budget. The principal clients are members of the People's Liberation Army, which consists of more than 2 million active personnel and more than 1 million reserves. Members of the army do not pay for consultations; civilians are charged approximately $US 100 per consultation. The network has its own website, a large satellite-based

communications system, more than 100 specialists and experts located in a central medical research center in Beijing, and a number of dedicated local hospitals and health centers.

### 4.7 SkyHealth (India)

Founded in 2008 and operated by World Health Partners, a nongovernmental organization based in the United States, SkyHealth is a telemedicine franchise program focusing on women. Approximately 1,200 SkyCare centers, run by female entrepreneurs to provide medical services in the villages in their area, make up the core of the network. These centers are linked to a panel of specialists in Delhi who can either refer patients to clinics or arrange transportation to a hospital. In addition to the SkyCare centers, there are about 120 telemedicine provision centers, 16 healthcare centers, 9 diagnostic clinics, and 1,400 rural pharmacies in the network.

SkyHealth focuses on women's health in rural India, with particular emphasis on family planning. The primary value proposition for the patient is contact with a trusted local female health provider who has access to specialized knowledge. SkyHealth also makes a distinct value proposition to the female entrepreneurs who run the centers: providing them with an independent source of income.

SkyHealth offers training and low-cost mobile solutions to franchised, local, non-specialist, female health providers in rural communities, which allows them to carry out teleconsultations, make diagnoses, prescribe treatments, and make referrals. It also has a database that franchisees can use to transmit and collect patient data. World Health Partners provides administrative support. A franchise costs about $US 3,000; patients are charged about $US 0.90 per consultation, part of which is used to support other parts of the network. World Health Partners and SkyCare centers are also supported by charitable donations, including contributions from the Bill and Melinda Gates Foundation.

### 4.8 Tactive Telemedicine (The Netherlands)

Alcoholism costs the Netherlands an estimated 2.58 billion euros per year. Developed by the Dutch company Tactus Addiction, Tactive Telemedicine provides online treatment for people suffering from alcohol addiction in The Netherlands. It allows structured asynchronous interaction between counselor and patient, while aid workers from recognized mental health services provide treatments on a franchise basis. The goal is to replicate one-to-one cognitive behavioral therapy with a professional counselor in an online environment. Tactive delivers more than 5,500 units of treatments per year, with most ranging from 12 to 16 weeks.

Most clients are individuals who are financed by a health insurance company or an employer. The principal offer for individuals is a dedicated, confidential service to help them overcome their alcohol addiction. Most users consider the anonymity of asynchronous counseling an additional advantage. For Dutch society, it is a cost-effective approach for dealing with alcoholism. Tactive also provides support for research into addictive behaviors.

Designed and maintained by Tactus Addiction, the IT platform is a relatively straightforward asynchronous communications system with a self-service website that hosts online forums. Between 20 and 25 trained counselors are employed on a full-time basis to provide advice and support. Clients pay a fee for the for-profit service. The initial set-up costs were partly covered by a grant from the Dutch government, and local and regional governments pay for statistics reports on the levels of alcohol addiction in their areas. The system is now being sold to other countries.

### 4.9 TeleMed-Escape (Italy)

TeleMed-Escape is an electronic managing system that sends digitally signed test results, which are accepted as medically and legally valid, directly to patients and doctors, either via a computer using Postesalute (the e-health unit of Poste Italiane, the Italian postal service) or on paper via Postel (Poste Italiane's print-and deliver service). The local health authority number 9 (ULSS No. 9) of Treviso, in Italy's Veneto region, developed the system. This authority provides healthcare services to 407,000 citizens in 37 municipalities. The main hospital is based in Treviso with a smaller unit in Oderzo. Between them, they have 1,272 hospital beds and 70 wards.

The clients of the system are the citizens of the Veneto region and the organizations that provide them with healthcare. The patient is able to get access to test results more quickly and without the inconvenience of having to visit the hospital to retrieve them. For the doctors, hospitals, and pharmacies involved, the system means fewer errors, faster turnaround times, and lower administrative costs. Clerical errors have been reduced by an estimated 10 percent. The system is funded and maintained by the local health authority and is currently free of charge. The initial outlay, primarily spent on systems integration rather than system development, was quickly recovered, thanks to the savings that have been generated.

### 4.10 Telenor TeleDoctor (Pakistan)

The Telenor Telecommunications Company, which is owned by the Norwegian Telenor Group, started the TeleDoctor program in Sindh, Pakistan, in 2008. It is now the second largest telemedicine service provider in Pakistan, with more than 26.1 million subscribers. It allows anyone in the network to connect to a doctor by dialing a single number. In effect, it is a sponsored hotline that dispenses non-specific medical advice and promotes the discussion of symptoms, treatments, and general health awareness. The service is available 24 hours a day, seven days a week, in eight languages.

The clients are all users of the Telenor mobile telephone network in Pakistan; the network itself also uses the service as a means to build customer loyalty and brand recognition. The primary value proposition for the patient is the ability to find a male or female doctor who speaks one of the eight languages used in Pakistan, at any time of the day or night. As with similar programs in India and China, it also offers ease of access and reduces travel costs, particularly in rural areas. Finally, for the participating doctors, it offers an additional source of income; for Telenor Communications, it is a way of building brand recognition.

The key resources are Telenor's mobile telephone network and the doctors that staff the helplines. Telenor uses the network both as a means of reaching the client and as a way to advertise the service. The cost of making a call is, $US0.08 per minute, which is used to offset the doctors' fees and the costs associated with marketing.

# Exhibit 3. Summary of the Case Studies

| Case Study | Location | Type | Offer | Client | Resources | Finance | URL |
|---|---|---|---|---|---|---|---|
| Apollo Telemedicine | India | Teleconsultation, education | Reduced travel costs, access to specialized knowledge for patient | Organizations and lower to middle-income individuals in rural India | IT platform, healthcare specialists, healthcare centers | Patient pays for consultation, government, and company pay for network. | http://www.apollotelehealth.com/ |
| Aravind Tele-Ophthalmology | India | Teleconsultation | Accessible and affordable eye care for patient | Low income and marginalized individuals in rural India | IT platform, volunteers, specialist telemedicine equipment | Patient pays for consultation | http://www.aravind.org/ |
| Arizona Telemedicine Program | United States | Application service provider | Reduce equipment, communication and network costs for institutions | Hospitals, prisons, and schools | Communications network, centralized database, purchasing power; expertise in training and consulting | Annual membership plus charge per service provided, Arizona state government provides grant | http://telemedicine.arizona.edu/ |
| Centro Unico di Prenotazione | Italy | e-Health | Reduced travel costs for patients, efficiency and flexibility for all; additional revenue for laboratories and pharmacies | Doctors, patients, laboratories and pharmacies | IT platform, database, network of doctors, laboratories and pharmacies | Payment of doctors, laboratories and pharmacies is via systems maintained by Umbrian regional government | http://www.cupsanita.it/ |
| Myca Nutrition | Canada | Teleconsultation | Convenient and timely method of monitoring of diet | Nutritionists and doctors | Internet and cell phone network | Monthly subscriptions | http://www.mycanutrition.com/ |

| Case Study | Location | Type | Offer | Client | Resources | Finance | URL |
|---|---|---|---|---|---|---|---|
| People's Liberation Army telemedicine network | China | Teleconsultation, e-health, education | Reduced travel costs, access to specialized knowledge for patient | Military personnel, and civilians in rural China | Satellite communications, military hospitals, and online library and database | Civilian patients pay for consultation; military patients paid for by government | http://www.301hospital.com.cn/en2012/web/Telemedicine.html |
| SkyHealth Network | India | Teleconsultation service provider | Reduced travel costs for patients, income from franchise for entrepreneur | Individuals in rural India | IT platform, specialist telemedicine equipment, entrepreneurs, rural health partners, doctors, and pharmacists | Patients pay for consultation; support from World Health Partners and the Bill and Melinda Gates Foundation | http://worldhealthpartners.org/ |
| Tactive Telemedicine | Netherlands | Teleconsultation | Confidentiality; dedicated personal assistance for patients; dealing with social problem | Companies; individuals with addiction problems | IT platform, counsellors, and franchisees | Clients pay a fee; local and regional governments pay for statistics; the application is re-sold to other countries | https://www.tactus.nl/teksten/item/bekijk/id/98 |
| TeleMed-Escape | Italy | e-Health | Operational efficiency and reduced waiting times for documents | Doctors and patients | Postel and Postesalute (Poste Italiane's eHealth unit) | Funded by local health authority | http://www.ulss.tv.it/categorie/link-utili/ritiro-referti.html |
| Telenor TeleDoctor | Pakistan | Teleconsultation | Reduced travel costs, immediate availability, multilingual, confidential | Patients, mostly in rural areas of Pakistan and the network itself | Telenor's mobile telephone network, a multilingual panel of doctors | Clients pay per minute via call charges | http://www.telenor.com.pk/talkshawk-infotainment/talkshawk-social-services-1911 |

## 5 Business Models for e-Health

A look at each of the four principal components in Osterwalder and Pigneur's (2010) business model canvas highlights the lessons learned from these ten examples of innovative e-health systems.

### 5.1 The Client

With the exception of the Arizona Telemedicine Program, which serves organizations rather than individuals, there are two main client groups in the case studies. The first group concerns clients who are marginalized in some way, such as those who have little or no income, or those who live in rural areas, far away from centralized resources. The second group concerns clients who have problems, such alcohol addiction or dietary issues, which are not connected to their physical or economic status. Each profile shapes the value proposition and the rest of the business model.

### 5.2 The Offer

There are a number of different value propositions in the case studies, but in most cases they hinge on offers made to the first group, marginalized clients, such as those who live in rural areas far away from centralized resources. Initially, it might appear that this type of client will only be found in large, rapidly developing economies, such as India and China; however, geographically isolated populations exist even in more mature economies (Kimble, 2011). Looking specifically at the problems of an aging population, it is worth noting that isolation and marginalization do not stem solely from geographic location; elderly patients with disabling conditions in a metropolis can be just as isolated as someone who lives in a rural setting. Teleconsultations can be used to bring healthcare to those who, regardless of location, find it difficult to travel to a doctor or hospital, such as the disabled, the elderly, or those housed in institutions.

For the second client group, the issue has less to do with availability than with acceptability in the long term. The business models associated with this client group are perhaps more relevant to the management of chronic conditions, such as diabetes or heart disease. Controlling these conditions requires patients to monitor their condition regularly and, often, to change certain aspects of their lifestyle. The business models adopted by companies like Myca Nutrition, Tactive Telemedicine, and Telenor TeleDoctor may be more easily applied to e-health initiatives aimed at dealing with chronic conditions, where patients have a degree of choice about the way they manage their health issues, than dealing with the problems of social or economic isolation.

### 5.3 Resources

Given that IT is often put forward to address the problems of healthcare provision, it is interesting to note that in most of the case studies presented here, the IT platforms are either relatively simple, standard technologies or are merely additions to existing systems or networks. In most cases, the key resources are human, such as doctors, health workers, or counselors, or links to broader networks that can be used to leverage existing resources, such as universities, pharmacies, or volunteer groups.

A particularly good illustration of this is the TeleMed-Escape system, where the key to the success of the initiative was finding a way to produce an electronic document that conformed to Italy's existing legal and medical codes. Once this was found, much of the system development that followed consisted of little more than ensuring that the disparate parts of the system could communicate with each other. Similarly, the success of the Centro Unico di Prenotazione depended, to a large extent, on the involvement of large numbers of local pharmacies, which was achieved by offering them an additional payment for each booking that was made through the system.

To some extent, this supports the observation made by Broens et al. (2007) that once an application has moved beyond a small-scale pilot program, the technology becomes less important than the business model. The implication for dealing with the problems considered here is that, if the right business model can be found, it is possible to save money and improve efficiency simply by making a few relatively minor adaptations to existing systems, rather than embarking on major investments in IT, such as electronic health records (Kimble, 2014).

### 5.4 Finance

The evaluation of costs and benefits in e-heath and healthcare in general is a perennial problem involving a large number of different stakeholders and sometimes conflicting views on the value of a particular course of action. As Heeks notes, "The first difficulty is the subjectivity of evaluation: Viewed from different perspectives, one person's failure may be another's success" (Heeks, 2006, p. 126).

Some, such as Gamble et al. (2004), have approached this problem by focusing only on the cost and benefits to healthcare providers, with the goal of developing business models that can be supported without external funding. For example, they note that some telemedicine applications fail to deliver adequate value given their costs and, while they offer considerable value to rural patients, also burden the healthcare provider with unrecoverable costs.

Most of the business models described above are not straightforward for-profit businesses. Although most require some sort of fee from the user, they also rely, directly or indirectly, on funding from governments or charities. Similarly, a number of the cases use activities in one area, such as education, to support or subsidize activities in another.

The philosophical, moral, economic, and political arguments about the funding of healthcare are beyond the scope of this article; however, the prevailing view about the provision or, as some see it, the rationing of healthcare must be taken into account. As noted earlier, this poses particular problems for the European Union, where individual member states have different approaches to funding healthcare.

## 6 The Future of Sustainable Business Models for e-Health

The experiences of the ten successful e-health initiatives from around the world described above shed light on the enormity of the task of containing spiraling healthcare costs while effectively and compassionately managing chronic diseases in an aging population. Certainly, some aspects of these organizations' business models are relevant to a discussion of how best to provide healthcare in the 21st century. Others, however, seem to be tied to the particular social and economic conditions that created the difficulties that these healthcare organizations face. Although, as Hart and Christensen (2002) argue, such business models may travel well, the challenges of transforming a business model designed to deal with, say, women's health issues in rural India into one that addresses the problems of aging patients with chronic conditions in the Unites States or Europe are considerable.

Information technology and appropriate business models surely are part of the solution to effective and efficient delivery of healthcare. Ultimately, however, the key to overcoming any problems encountered along the way to achieving that goal may lie in the next step in Broens et al.'s (2007) layered implementation model: the legal and social environment in which these decisions take place.

# 7 References

Broens, T. H., Vollenbroek-Hutten, M. M., Hermens, H. J., van Halteren, A. T., & Nieuwenhuis, L. J. (2007). Determinants of successful telemedicine implementations: A literature study. Journal of telemedicine and telecare, 13(6), 303-309.

Celler, B. G., Lovell, N. H., & Basilakis, J. (2003). Using information technology to improve the management of chronic disease. Medical Journal of Australia, 179(5), 242-246.

Chen, S., Cheng, A., & Mehta, K. (2013). A review of telemedicine business models. Telemedicine and e-Health, 19(4), 287-297.

Chesbrough, H. W. (2010). Business model innovation: Opportunities and barriers. Long Range Planning, 43(2-3), 354-363.

Chesbrough, H. W., & Rosenbloom, R. S. (2002). The role of the business model in capturing value from innovation: Evidence from Xerox Corporation's technology spin-off companies. Industrial & Corporate Change, 11(3), 529-555.

Currie, W. L., & Seddon, J. J. M. (2014). A cross-national analysis of eHealth in the European Union: Some policy and research directions. Information & Management, 51(6), 783-797.

Davenport, T. H., & Glaser, J. (2002). Just-in-time delivery comes to knowledge management. Harvard Business Review, 80(7), 107 - 111.

EU. (2004). e-Health — Making healthcare better for European citizens: An action plan for a European e-health area. Brussels: Commission of the European Communities.

Eysenbach, G. (2001). What is e-health? Journal of Medical Internet Research, 3(2), e20.

Gamble, J. E., Savage, G. T., & Icenogle, M. L. (2004). Value-chain analysis of a rural health program: toward understanding the cost benefit of telemedicine applications. Hospital Topics, 82(1), 10-17.

Hart, S., & Christensen, C. M. (2002). The great leap: Driving innovation from the base of the pyramid. MIT Sloan Management Review, 44(1), 51-56.

Heeks, R. (2006). Health information systems: Failure, success and improvisation. International Journal of Medical Informatics, 75(2), 125-137.

Hill, J. W., & Powell, P. (2009). The national healthcare crisis: Is eHealth a key solution? Business Horizons, 52(3), 265-277.

Kimble, C. (2011). Building effective virtual teams: How to overcome the problems of trust and identity in virtual teams. Global Business and Organizational Excellence, 30(2), 6-15.

Kimble, C. (2014). Electronic health records: Cure-all or chronic condition? Global Business and Organizational Excellence, 33(4), 63-74.

Kimble, C., & Bourdon, I. (2013). The link among information technology, business models, and strategic breakthroughs: Examples from Amazon, Dell, and eBay. Global Business and Organizational Excellence, 33(1), 58-68.

Linder, J., & Cantrell, S. (2000). Changing business models: Surveying the landscape. Cambridge, Massachusetts: Accenture Institute for Strategic Change.

Mrsnik, M., & Morozov, I. (2012). Mounting medical care spending could be harmful to the G-20's credit health. New York: Standard & Poor's Financial Services.

Osterwalder, A., & Pigneur, Y. (2010). Business model generation: A handbook for visionaries, game changers, and challengers. Wiley.com.

Osterwalder, A., Pigneur, Y., & Tucci, C. L. (2005). Clarifying business models: Origins, present, and future of the concept. Communications of the Association for Information Systems, 16(1), 1 - 25.

Valeri, L., Giesen, D., Jansen, P., & Klokgieters, K. (2010). Business models for eHealth. Cambridge, UK: RAND Europe and Capgemini Consulting.

Zundel, K. M. (1996). Telemedicine: History, applications, and impact on librarianship. Bulletin of the Medical Library Association, 84(1), 71-79.